\documentclass[12pt,a4paper]{JHEP3}

\usepackage{amssymb, amsmath, amsopn, amsthm, dsfont}
\usepackage{epsfig}

\allowdisplaybreaks

\newcommand{\beq}{\begin{equation}}
\newcommand{\eeq}{\end{equation}}

\newcommand{\be}{\begin{equation}}
\newcommand{\ee}{\end{equation}}
\newcommand{\bea}{\begin{eqnarray}}
\newcommand{\eea}{\end{eqnarray}}
\newcommand{\bean}{\begin{eqnarray*}}
\newcommand{\eean}{\end{eqnarray*}}
\newcommand{\nn}{\nonumber}

\title{Higher spins in the symmetric orbifold of K3}

\author{Marco Baggio, Matthias R.\ Gaberdiel, and Cheng Peng\\

{\tt \{baggiom, gaberdiel, pengch\}@itp.phys.ethz.ch}\\
Institut f\"ur Theoretische Physik, ETH Zurich,
CH-8093 Z\"urich, Switzerland \\
}

\abstract{The symmetric orbifold of K3 is believed to be the CFT dual of string theory on ${\rm AdS}_3 \times {\rm S}^3 \times {\rm K3}$
at the tensionless point. For the case when the K3 is described by the orbifold $\mathbb{T}^4/\mathbb{Z}_2$, we identify a subsector of 
the symmetric orbifold theory that is dual to a higher spin theory on AdS$_3$. We analyse how the BPS spectrum of string theory 
can be described from the higher spin perspective, and determine which single-particle BPS states are 
accounted for by the perturbative higher spin theory.
}

\preprint{}

\allowdisplaybreaks

\begin{document}
\section{Introduction}
One of the long standing puzzles in theoretical physics is the determination of the symmetry algebra of string theory.  
Hints of an enlarged symmetry in string theory have been found by studying the high energy limit 
\cite{Gross:1988ue, Witten:1988zd, Moore:1993qe}, 
where a large set of higher spin symmetries emerge. In asymptotically AdS backgrounds, this limit can be studied in quite some 
detail thanks to the AdS/CFT correspondence \cite{Maldacena:1997re}, as it corresponds to free gauge theories on the boundary
\cite{Sundborg:2000wp,Mikhailov:2002bp,Sezgin:2002rt}. The 
extended higher-spin gauge symmetries of the bulk are then dual to the large number of conserved currents that 
emerge in the free limit of the dual theory. 
 
This idea has sparked considerable interest in studying this higher-spin subsector in isolation from the rest of the dynamics, 
which has led to a number of weak-weak 'vector-like' dualities between Vasiliev \cite{Vasiliev:2003ev}
higher-spin theories on AdS 
and CFTs with extended symmetries on the boundary \cite{Klebanov:2002ja,Sezgin:2003pt}.
  
In the AdS${}_3$/CFT${}_2$ context, the relevant higher-spin bulk theories have been conjectured to be dual to minimal model CFTs 
\cite{Gaberdiel:2010pz}. In this paper, we are particularly interested in theories with $\mathcal{N}=4$ supersymmetry
that arise as the near-horizon limit of D1-D5 systems; in this case the bulk geometry is 
AdS${}_3 \times {\rm S}^3 \times \mathcal{M}_4$, where $\mathcal{M}_4$ is either $\mathbb{T}^4$ or K3. In the tensionless limit
of string theory,  
these backgrounds are believed to be dual to the free symmetric orbifold $\mathrm{Sym}_N \mathcal{M}_4$, where $N=Q_1 Q_5$ is the 
product of the D1 and D5 brane charges. On the other hand, their higher-spin subsector should be dual to minimal models with $\mathcal{N}=4$ 
symmetry.

It is then natural to try and understand the relation between the minimal model CFTs dual to the higher-spin subsector 
of such theories,  and the  
CFTs (in the present case, the symmetric orbifold theory) describing the full string theory spectrum. This has been answered in 
\cite{Gaberdiel:2014cha} for the case of AdS${}_3 \times {\rm S}^3 \times \mathbb{T}^4$ in the limit where the volume of the torus 
is very large. In this limit, the perturbative part of the Vasiliev AdS${}_3$ theory is captured by a subsector of the large level limit of 
particular Wolf space cosets \cite{Schoutens:1988ig,Spindel:1988sr,Goddard:1988wv,VanProeyen:1989me,Sevrin:1989ce}.\footnote{
This subsector is not modular invariant by itself and needs to be completed with additional states. 
The usual diagonal modular invariant leads to a plethora of light states, see e.g.~\cite{Gaberdiel:2013cca}, whose precise
interpretation from a bulk viewpoint has not been understood in detail, see however
\cite{Castro:2011iw,Perlmutter:2012ds}.}  
Indeed, it was shown in \cite{Gaberdiel:2014cha} that the relevant subsector is described by the ${\rm U}(N-1)$
invariant states of the $4N$ free bosons and fermions that make up the $\mathbb{T}^{4N}$ theory of the symmetric orbifold;
this is naturally a subsector of the $S_N$ invariant states of the symmetric orbifold since $S_N\subset {\rm U}(N-1)$. 
The symmetric orbifold can then be regarded as another modular invariant of the coset theory. All its states can
be organised in terms of the $\mathcal{W}$-algebra associated to the coset CFT, but it does not possess any light states.

In this paper we want to find the analogous relation for the case where ${\cal M}_4={\rm K3}$. We shall concentrate on
a specific K3 sigma-model, namely the one where ${\rm K3}$ can be described by the orbifold ${\rm K3} = \mathbb{T}^4 / \mathbb{Z}_2$. 
The situation is then a little different from what was considered in \cite{Gaberdiel:2014cha} since the
symmetric orbifold of K3 does not actually contain a contraction of the Wolf-space large ${\cal N}=4$ ${\cal W}_{\infty}[0]$ algebras 
as a symmetry algebra --- this
follows, for example, from the fact that the elliptic genus of the symmetric orbifold of K3 does not vanish, while that of the
large ${\cal N}=4$ theories is always trivial. The higher spin -- CFT duality that is relevant for the K3 case is therefore slightly
different from what was considered before in \cite{Gaberdiel:2013vva}: at $\lambda=0$ both the higher spin algebra
and the Wolf space cosets contain a subtheory, and it is the duality between these subtheories that is of relevance here. 
In particular, we show that the full spectrum of the symmetric K3-orbifold can be organised in terms of representations
of an appropriate subalgebra ${\cal W}_\infty^s$ of ${\cal W}_\infty[0]$. 

One added benefit of repeating the analysis for K3 (rather than $\mathbb{T}^4$) is that the elliptic genus of the
K3 symmetric orbifold is non-trivial. Thus it is natural to look at the BPS spectrum of the theory, and we 
analyse the BPS spectrum of supergravity or string theory on AdS${}_3 \times {\rm S}^3 \times {\rm K3}$
in terms of the representation theory of ${\cal W}_\infty^s$. This allows us to determine the part of the BPS spectrum that is captured by the
perturbative higher spin theory. As it turns out, only a tiny fraction of the single-particle BPS states of string theory
are actually contained in the perturbative higher spin theory; for example, out of the $21$ chiral primaries whose 
descendant are the exactly marginal operators
of the K3 symmetric orbifold, only $2$ are part of the perturbative higher spin theory. 
\smallskip

The paper is organised as follows. In Section~2, we introduce the modified higher spin -- CFT duality that is relevant in our
case, and we explain how it is related to the symmetric orbifold of K3. In Section~3 we  show that the chiral algebra
of the K3 symmetric orbifold can indeed be described in terms of representations of the relevant ${\cal W}_\infty^s$ algebra.
In Section~4 we study the chiral primaries of the symmetric orbifold from the higher spin perspective; we identify all 
low-lying states of the symmetric orbifold in terms of  ${\cal W}_\infty^s$ representations, and then specifically identify the 
BPS states among them. We also enumerate the BPS states of the perturbative higher spin theory and show
that there are only two single-particle states among them. Finally, Section~5 contains our conclusions.

\section{The small $\mathcal{N}=4$ higher-spin duality}

Let us begin by reviewing the minimal model holography with large $\mathcal{N}=4$ superconformal symmetry. The relevant
higher spin theory is based on the Lie superalgebra ${\rm shs}_{2}[\lambda]$, and was argued to be dual to the Wolf space coset
CFTs \cite{Gaberdiel:2013vva}
\begin{equation}
	\label{eq:coset}
	\frac{\mathfrak{su}(N+2)^{(1)}_{k+N+2}}{\mathfrak{su}(N)^{(1)}_{k+N+2}\oplus \mathfrak{u}(1)^{(1)}_{\kappa}}\oplus \mathfrak{u}(1)\ , 
\end{equation}
where $\kappa = 2N(N+2)(N+k+2)$. These cosets contain the large $\mathcal{N}=4$ superconformal algebra 
$A_\lambda$ with $\lambda = \frac{N+1}{N+k+2}$ (the reader is encouraged to consult \cite{Gaberdiel:2013vva} for more details);
further evidence for this duality was given in \cite{Creutzig:2013tja,Candu:2013fta,Gaberdiel:2014yla,Beccaria:2014jra,Candu:2014yva}. 

Except
for very small values of $N$ and $k$, the chiral algebra contains higher spin currents in addition to the ones associated to the
superconformal symmetry. These can be organised into multiplets $R^{(s)}$ of the superconformal algebra, where 
$s=1,2,\ldots$ labels the spin of the highest-weight state,
\begin{equation}
\begin{array}{lcc}\label{D2mult2}
& s: & ({\bf 1},{\bf 1})  \\
& s+\tfrac{1}{2}: & ({\bf 2},{\bf 2})  \\
R^{(s)}: \qquad & s+1: & ({\bf 3},{\bf 1}) \oplus ({\bf 1},{\bf 3})  \\
& s+\tfrac{3}{2}: & ({\bf 2},{\bf 2})  \\
& s+2: & \  \  ({\bf 1},{\bf 1})  \ .
\end{array}
\end{equation}
The quantum numbers shown in the column on the right refer to the two $\mathfrak{su}(2)$ algebras of $A_\lambda$. Together with the 
superconformal algebra generators, they form the large $\mathcal{N}=4$ $\mathcal{W}_\infty[\lambda]$ algebra, which is studied in detail in 
\cite{Gaberdiel:2014yla,Beccaria:2014jra}. 

In order to obtain a higher-spin algebra with small $\mathcal{N}=4$ symmetry, we now take the limit 
$k\rightarrow \infty$, for which $\lambda\rightarrow 0$. In this limit, the 
large $\mathcal{N}=4$ algebra contracts to the small $\mathcal{N}=4$ algebra together with $4$ free bosons and $4$ free fermions. 
In terms of the coset description, taking the limit $k \to \infty$ leads to the continuous orbifold of the form 
\cite{Gaberdiel:2014cha} (see also \cite{Gaberdiel:2011aa,Gaberdiel:2014vca})
\begin{equation}
	\label{eq:contorb1}
	(\mathbb{T}^4)^{N+1} \Big/ \mathrm{U}(N)~.
\end{equation}
There are also matter fields corresponding to the 
$(0;\mathrm{f})\otimes \overline{(0;\mathrm{f}^*)}$  and $(0;\mathrm{f}^*)\otimes \overline{(0;\mathrm{f})}$ degrees of freedom of the CFT, 
and the perturbative part of the bulk higher spin theory is described by the CFT subsector
\begin{equation}
	\label{eq:hpert}
	\mathcal{H}^\mathrm{(pert)} = \bigoplus_{\Lambda}\,  (0;\Lambda)\otimes\overline{(0;\Lambda^*)}\ .
\end{equation}
The authors of \cite{Gaberdiel:2014cha} found an interesting non-diagonal modular completion of the above Hilbert space, 
which corresponds to the degrees of freedom of string theory on AdS${}_3 \times {\rm S}^3 \times \mathbb{T}^4$ in the tensionless limit --- the 
dual CFT can then be described in terms of the symmetric orbifold $\mathrm{Sym}_{N+1}(\mathbb{T}^4)$, all of its states can be 
organised into representations of the $\mathcal{W}_\infty[0]$ algebra.

It is natural to expect that a similar analysis could be carried out for string theory on AdS${}_3 \times {\rm S}^3 \times \mathrm{K3}$, 
since the dual theory also enjoys small $\mathcal{N}=4$ supersymmetry. However, we find a small difficulty in this: the chiral algebra of 
$\mathrm{Sym}_{N}(\mathrm{K3})$ does \emph{not} contain the $\mathcal{W}_\infty[0]$ algebra described above as a subalgebra. This is 
due to the fact that the supersymmetry algebra of K3 is not a contracted version of the large $\mathcal{N}=4$ algebra, as can be easily 
seen for example by noticing that the elliptic genus of K3 does not vanish. However, the chiral algebra at the symmetric orbifold point 
does contain the $\mathcal{W}^s_\infty$ algebra, which is obtained from $\mathcal{W}_\infty[0]$ by removing the 4 free bosons 
and fermions, and for $\lambda=0$, the resulting coset theory can be described as (see Section~\ref{sec:2.1} below)
\begin{equation}
	\label{eq:t4NmodUN}
	(\mathbb{T}^4)^{N} \Big/ \mathrm{U}(N)\ .
\end{equation}
Note that now the superconformal algebra contains only one $\mathfrak{su}(2)$ algebra, which is nothing else but the R-symmetry 
algebra of the small $\mathcal{N}=4$ superconformal algebra.

Analogously, the bulk theory can be based on the higher-spin algebra shs${}^s_2$, which is obtained from shs${}_2[0]$ upon
removing the generators 
\begin{equation}
( 1 + k) \otimes E_{\alpha\beta} \ , 
\end{equation}
which can never appear in commutators, see \cite{Gaberdiel:2013vva} for our conventions.\footnote{At $\lambda=0$, $\nu=-1$, and hence the
commutator $[\hat{y}_\alpha,\hat{y}_\beta] = 2 i \epsilon_{\alpha\beta} (1 - k)$ is proportional to $(1-k)$.} Since the representation theory of the 
$\mathcal{W}^s_\infty$ algebra is largely identical to that of $\mathcal{W}_\infty[0]$ --- in particular, representations of $\mathcal{W}^s_\infty$
can also be labeled by pairs $(\Lambda_+;\Lambda_-)$, and the wedge characters agree --- it is immediate that we can deduce a 
higher spin/CFT correspondence 
of the form
\begin{equation}\label{dualitys}
\hbox{higher spin theory based on  shs${}^s_2$} \ \longleftrightarrow \ (\mathbb{T}^4)^{N} \Big/ \mathrm{U}(N)\ .
\end{equation}
We want to show in this paper that string theory on ${\rm AdS}_3 \times {\rm S}^3 \times {\rm K3}$ can be interpreted within this framework.

\subsection{The continuous orbifold and the symmetric orbifold}\label{sec:2.1}

It was shown in \cite{Gaberdiel:2014cha}, see also \cite{Gaberdiel:2014vca}, that the 
large $k$ limit of the cosets \eqref{eq:coset} can be described by the continuous orbifold \eqref{eq:contorb1}. 
In this limit, removing the 4 free bosons and fermions corresponds to removing  the `center-of-mass' 
$\mathbb{T}^4$ in \eqref{eq:contorb1}, and the 
remaining fields transform as 
\begin{eqnarray}\label{bosfertrans0}
\hbox{bosons:} & \qquad & 2 \cdot ({\bf N},{\bf 1}) \oplus  2 \cdot (\overline{\bf N},{\bf 1}) \nonumber \\
\hbox{fermions:} & \qquad & ({\bf N},{\bf 2}) \oplus (\overline{\bf N},{\bf 2}) \ , 
\end{eqnarray}
where the labels refer to the representations of the U($N$) and SU(2) R-symmetry, respectively. The complex
boson field of the higher spin theory corresponds to the representations 
$(0;\mathrm{f})\otimes \overline{(0;\mathrm{f}^*)}$  and $(0;\mathrm{f}^*)\otimes \overline{(0;\mathrm{f})}$ of the dual
CFT. As in the duality considered in \cite{Gaberdiel:2014cha}, 
the perturbative part of the higher spin theory based on ${\rm shs}_2^s$ is thus captured by the states of the form
$(0;\Lambda) \otimes \overline{(0;\Lambda^\ast)}$, where $\Lambda$ runs over the ${\rm U}(N)$ representations with finitely many
boxes and anti-boxes, i.e., by eq.~(\ref{eq:hpert}).

We would like to relate this continuous orbifold to the symmetric orbifold describing string theory on 
AdS${}_3 \times {\rm S}^3 \times \mathrm{K3}$ in the tensionless limit, namely
\begin{equation}
\label{eq:k3modsn}
	\mathrm{Sym}_{N}(\mathrm{K3}) \equiv (\mathrm{K3})^N \Big/ S_N\,.
\end{equation}
Furthermore, it is easiest to establish the relation between the two theories at a particular point of the moduli space of K3, 
namely the orbifold point
\begin{align}
	\mathrm{K3} = \mathbb{T}^4 \Big/ \mathbb{Z}_2\ .
\end{align}
For then, the symmetric orbifold in \eqref{eq:k3modsn} can be written as 
\begin{align}
	\label{eq:t4modsnzn}
	\mathrm{Sym}_{N}(\mathrm{K3}) = (\mathbb{T}^4)^N \Big/ (S_{N}\ltimes \mathbb{Z}_2^{N})\,,
\end{align}
where the group $S_{N}\ltimes \mathbb{Z}_2^{N}$ is the semidirect product of the symmetric group $S_N$ and $N$ copies of $\mathbb{Z}_2$, 
and $S_N$ acts on $\mathbb{Z}_2^N$ by permuting the factors in the obvious way. The untwisted sector of this theory consists of $4N$ free 
bosons and fermions that transform as
\begin{eqnarray}\label{bosfertrans1}
\hbox{bosons:} & \qquad & 4 \cdot (N,{\bf 1}) \nonumber \\
\hbox{fermions:} & \qquad & 2\cdot (N,{\bf 2}) 
\end{eqnarray}
with respect to $S_{N}\ltimes \mathbb{Z}_2^{N} \times \mathrm{SU}(2)$, where SU(2) is again the R-symmetry of the 
small $\mathcal{N}=4$ algebra. Here $N$ denotes the $N$-dimensional representation of $S_{N}\ltimes \mathbb{Z}_2^{N}$, 
where the permutation group acts in the usual way (by matrices with one $1$ in each row and column), while
$\mathbb{Z}_2^{N}$ is described by the diagonal matrices with $\pm 1$ along the diagonal. We should
stress that this representation is an \emph{irreducible} representation of 
$S_{N}\ltimes \mathbb{Z}_2^{N}$. One way to see this is to note that, as a representation of $S_N$, it decomposes as 
$N \cong 1 + (N-1)$, 
where the $1$ is generated by the sum of all $N$ basis vectors. However, since this vector is not invariant under
the action of $\mathbb{Z}_2^{N}$, there is no non-trivial invariant subspace, and the representation is irreducible.
\smallskip

For the following it will be important that we have the obvious embedding
\begin{align}
	\label{eq:branching}
	S_{N}\ltimes \mathbb{Z}_2^{N} \hookrightarrow \mathrm{U}(N)~.
\end{align}
Under this embedding, both the ${\bf N}$ and the $\overline{\bf N}$ representations of $\mathrm{U}(N)$ branch into the 
$N$ of $S_{N}\ltimes \mathbb{Z}_2^{N}$. From this, and comparing \eqref{bosfertrans0} with \eqref{bosfertrans1}, we 
conclude that the untwisted sector of \eqref{eq:t4NmodUN} is a subsector of the untwisted sector of \eqref{eq:t4modsnzn}. 
As in the case of $\mathbb{T}^4$, the partition function of the symmetric orbifold provides then a non-diagonal modular 
invariant completion for \eqref{eq:hpert}.

\section{Chiral algebra}
In this section, we decompose the chiral algebra of the symmetric orbifold of K3 in terms of representations of the small $\mathcal{W}^s_\infty$ algebra described in the previous section.

\subsection{The chiral algebra of $\mathrm{Sym}_{N}(\mathrm{K3})$}

We want to show that, for sufficiently large $N$, the vacuum character $\mathcal{Z}_{\rm vac}(q,y)$ of 
$\mathrm{Sym}_{N}(\mathrm{K3})$ can be decomposed as
\begin{equation}
	\mathcal{Z}_{\rm vac}(q,y) = \sum_{\Lambda} n(\Lambda) \, \chi_{(0;\Lambda)} (q,y)\ , 
\end{equation}
where $\chi_{(0;\Lambda)}$ are the $\mathcal{W}^s_\infty$ characters of the representations $(0;\Lambda)$, which in turn belong to the 
untwisted sector of the continuous orbifold \eqref{eq:t4NmodUN}. The non-negative integers $n(\Lambda)$ denote the multiplicity of the trivial 
representation of $S_{N}\ltimes \mathbb{Z}_2^{N}$ in the branching of the $\mathrm{U}(N)$ representation 
$\Lambda$ under \eqref{eq:branching}; they can be computed using standard character techniques. 

The vacuum character of $\mathrm{Sym}_{N}(\mathrm{K3})$ can be easily deduced from the vacuum character of K3 by 
using the DMVV formula \cite{Dijkgraaf:1996xw}. In fact, if we write
\begin{equation}
	\mathcal{Z}^{\rm chiral}_{\rm R}(\mathrm{K3}) = \mathrm{Tr}_{R} (-1)^F q^{L_0} y^{J_0} = \sum_{\Delta,\ell} c(\Delta,\ell) \,
	q^\Delta y^\ell\ ,
\end{equation}
the corresponding chiral character of the symmetric orbifold is given by
\begin{equation}
	\sum_{k=0}^\infty \mathcal{Z}^{\rm chiral}_{\rm R}(\mathrm{Sym}_{N}(\mathrm{K3})) = \prod_{\Delta,\ell} 
	\frac{1}{(1-pq^\Delta y^\ell)^{c(\Delta,\ell)}}\ .
\end{equation}
Here the subscript R refers to the fact that we are working in the Ramond sector. It is then easy to recover the chiral character in the 
NS sector by spectral flow. For sufficiently large $N$, the result is (we suppress the overall factor of $q^{-N/4}$)
\begin{align}
\nonumber \mathcal{Z}_{\rm vac} &=1+q \left(y^2+\frac{1}{y^2}+4\right)+q^{3/2} \left(8 y+\frac{8}{y}\right)\\
\nonumber &+q^2 \left(y^4+\frac{1}{y^4}+8 y^2+\frac{8}{y^2}+30\right)+q^{5/2} \left(8 y^3+\frac{8}{y^3}+64 y+\frac{64}{y}\right)\\
&+q^3 \left(y^6+\frac{1}{y^6}+8 y^4+\frac{8}{y^4}+93 y^2+\frac{93}{y^2}+248\right)+\mathcal{O}(q^{7/2})\ .\label{extw}
\end{align}
In analogy with \cite{Gaberdiel:2014cha}, we can decompose this character into coset representations as
\begin{align}
\nonumber \mathcal{Z}_{\rm vac}&=\chi_{ (0, [0,0,\ldots,0,0])}+\chi_{ (0, [2,0,\ldots,0,0])}+\chi_{ (0, [0,0,\ldots,0,2])}+2 \chi_{ (0, [4,0,\ldots,0,0])}+2 \chi_{ (0, [0,0,\ldots,0,4])}\\
\nonumber &+\chi_{ (0, [0,2,0,\ldots,0,0])}+\chi_{ (0, [0,0,\ldots,0,2,0])}
+\chi_{ (0, [3,0,\ldots,0,1])}+\chi_{ (0, [1,0,\ldots,0,3])}+2 \chi_{ (0, [2,0,\ldots,0,2])}\\
\nonumber & +3 \chi_{ (0, [6,0,\ldots,0,0])}+3 \chi_{ (0, [0,0,\ldots,0,6])}+\chi_{ (0, [4,1,0,\ldots,0,0])}+\chi_{ (0, [0,0,\ldots,0,1,4])}\\
\nonumber &+2 \chi_{ (0, [2,2,0,\ldots,0,0])} +2 \chi_{ (0, [0,0,\ldots,0,2,2])}+\chi_{ (0, [0,0,2,0,\ldots,0,0])}+\chi_{ (0, [0,0,\ldots,0,2,0,0])} \\
\nonumber &+2 \chi_{ (0, [5,0,\ldots,0,1])}+2 \chi_{ (0, [1,0,\ldots,0,5])}+\chi_{ (0, [3,1,0,\ldots,0,1])}+\chi_{ (0, [1,0,\ldots,0,1,3])}\\
\nonumber&+\chi_{ (0, [1,2,0,\ldots,0,1])}+\chi_{ (0, [1,0,\ldots,0,2,1])}+4 \chi_{ (0, [4,0,\ldots,0,2])}+4 \chi_{ (0, [2,0,\ldots,0,4])}\\
\nonumber &+\chi_{ (0, [2,1,0,\ldots,0,2])}+\chi_{ (0, [2,0,\ldots,0,1,2])}+2 \chi_{ (0, [0,2,0,\ldots,0,2])}+2 \chi_{ (0, [2,0,\ldots,0,2,0])}\\
 &+3 \chi_{ (0, [3,0,\ldots,0,3])}+\chi_{ (0, [3,0,\ldots,0,1,1])}+\chi_{ (0, [1,1,0,\ldots,0,3])}+\mathcal{O}(q^4)\ , \label{vacch}
\end{align}
and the multiplicities agree with the group theoretic prediction for $n(\Lambda)$ from above. Here the wedge characters of the
representations $(0;\Lambda)$ agree with those given in \cite{Gaberdiel:2014cha}, and the only difference concerns the contribution
of the modes outside the wedge --- these differ from \cite{Gaberdiel:2014cha} by the absence of the $4$ free bosons and fermions,
i.e., the free fermion and free boson contributions in eq.~(B.5) of \cite{Gaberdiel:2014cha}.

\subsection{The generating fields of the chiral algebra}
In order to compare the extended chiral algebra we just found to the coset $\mathcal{W}$-algebra, it is useful to 
identify its independent generators. The small $\mathcal{N}=4$ $\mathcal{W}^s_\infty$ algebra is generated by 
$8$ fields for each (half-integer) spin greater than 1, as well as $4$ fields at $s=1$.
This information can be conveniently encoded in the generating function $J(q,y)$, defined by
\begin{align}
\label{k3red}
J(q,y)& = q \left(y^2+\frac{1}{y^2}+2\right)+ \frac{q^{3/2}}{1-q} \left(4 y+\frac{4}{y}\right)+ \frac{q^2}{1-q} \left(y^2+\frac{1}{y^2}+6\right)\ , 
\end{align}
where the power of $y$ keeps track of the ${\rm U}(1)$ charge of the corresponding generator.
The analogous generating function for the extended chiral algebra of the K3 symmetric orbifold, which counts the independent fields 
at each dimension, can be calculated by removing all the contributions from descendants and products of lower level states from 
\eqref{extw}, and we get
\begin{align}
\nn J^{\text{K3}}(q,y)&= q \left(y^2+\frac{1}{y^2}+4\right)+q^{3/2} \left(8 y+\frac{8}{y}\right)+q^2 \left(3 y^2+\frac{3}{y^2}+15\right)\\
&+q^{5/2} \left(16 y+\frac{16}{y}\right)+q^3 \left(15 y^2+\frac{15}{y^2}+57\right)+\mathcal{O}(q^{7/2})\ .\label{k3red2}
\end{align}
As in the case of \cite{Gaberdiel:2015mra}, see eq.~(2.9), this generating function is related to the character of a single 
copy of K3 as
\begin{align}
J^{\rm K3}&=(1-q)\, \bigl(\mathcal{Z}^{\rm chiral}_{\rm NS}({\rm K3}) - 1 \bigr) \ ,
\end{align}
where $\mathcal{Z}^{\rm chiral}_{\rm NS}({\rm K3})$ is the untwisted chiral partition function of one copy of K3 in the NS sector.  
Following the same logic as in \cite{Gaberdiel:2015mra}, we have checked that we can organise the generating function 
\eqref{k3red2} in terms of coset representations as
\begin{align}
J^{\text{K3}}(q,y)&=J(q,y)+ (1-q)\, \sum_{m,n\geq 0}{}' \, \chi_{(0;[m,0\ldots 0,n])}(q,y)\ ,\label{k3rel}
\end{align}
where the prime on the summation symbol means that we sum over $m$ and $n$ for which $m+n$ is even,
and we exclude the cases $(m,n)=(0,0)$ and $(1,1)$. This describes rather succinctly the additional 
generating fields that need to be added to the small $\mathcal{N} = 4$ $\mathcal{W}^s_\infty$-algebra in order to obtain 
the extended chiral algebra of \eqref{eq:k3modsn}.

\section{Chiral primaries}
In this section we shall study more general states of the symmetric orbifold that do not belong to the chiral algebra. We shall focus
on the chiral primaries, and specifically on those of dimension $(h,\bar h) = (1/2,1/2)$ since their descendants contain
singlet states of the R-symmetry with $(h,\bar h) = (1,1)$  that describe exactly marginal deformations 
preserving the superconformal algebra. We will see that states with $h=\bar{h}=1/2$  arise from the untwisted as well as the 
$(12)$-twisted sectors of the symmetric orbifold, and we will identify their $\mathcal{W}$-algebra representations. We will
also study how many of the BPS states (and in particular of the $21$ chiral primaries associated to the exactly marginal deformations of the K3 symmetric orbifold) 
are accounted for in the perturbative Vasiliev theory.

\subsection{Chiral primaries in $\mathcal{N} = (4,4)$ theories}

Let us begin by reviewing the BPS spectrum of ${\cal N}=(4,4)$ theories. 
The representations of the small $\mathcal{N} = (4,4)$ algebra are characterised by the conformal dimension $(h,\bar h)$ as well as the 
R-symmetry quantum numbers $(j,\bar \jmath)$ under the left and right SU(2)${}_{\rm R}$. In our conventions, the unitarity bound is
\begin{equation}
h \geq \frac{j}{2}\ , \qquad \bar h \geq \frac{\bar \jmath}{2} \ .
\end{equation}
Representations that saturate any of the two bounds are shorter than generic representations, those that saturate both bounds are 
called \emph{chiral primaries}. As a consequence, these representations are characterised by the two SU(2) quantum numbers as 
$(j,\bar \jmath)_S$, where $S$ stands for `short'. A particularly interesting set of chiral states are those of the form $(1,1)_S$. These short 
multiplets contain four (descendant) states with $(h,\bar h)=(1,1)$ and $(j,\bar \jmath) = (0,0)$, that describe exactly marginal deformations 
preserving the small $\mathcal{N}=4$ superconformal algebra.

A very useful property of $\mathcal{N} = (4,4)$ (as well as $\mathcal{N} = (2,2)$)  theories in 2d is that the spectrum of chiral 
primaries is bounded from above. More precisely we have \cite{Lerche:1989uy}
\begin{equation}
	h \leq \frac{c}{6}\ .
\end{equation}
The spectrum of these theories can then be encoded in the so-called generalised Poincar\'e polynomial, defined as
\begin{equation}
	P_{t,\bar t} = \mathrm{Tr}\, t^{J_0} \bar t^{\bar J_0}\ ,
\end{equation}
where the trace is taken over the chiral primaries only. For supersymmetric sigma models with target space $\mathcal{M}$, the 
Poincar\'e polynomial is given by \cite{Witten:1981nf}
\begin{equation}
P_{t,\bar t} = \sum_{p,q} h^{p,q} \, t^p\, \bar t^q\ ,
\end{equation}
where the $h^{p,q}$ are the Betti numbers of $\mathcal{M}$. For symmetric orbifolds $\mathcal{M}^N / S_N$, the Poincar\'e polynomial 
can be computed from the formula \cite{gottsche1993perverse,deBoer:1998ip}
\begin{equation}
\sum_{N} Q^N P_{t,\bar t}(\mathcal{M}^N / S_N) = \prod_{m=1}^\infty \prod_{p,q} (1+(-1)^{p+q+1}Q^m t^{p+\frac{d}{2}(m-1)} \bar t^{q+\frac{d}{2}(m-1)})^{(-1)^{p+q+1} h^{p,q}}\ ,
\end{equation}
where $d$ is the complex dimension of $\mathcal{M}$.

We now specialise the discussion above to the case at hand where $\mathcal{M} = \mathrm{K3}$. The Betti numbers can be read off
from the Hodge diamond
\begin{equation}
\begin{array}{ccccc}
&&\phantom{1}1\phantom{1}&&\\
&\phantom{1}0\phantom{1}&&\phantom{1}0\phantom{1}&\\
\phantom{1}1\phantom{1}&&20&&\phantom{1}1\phantom{1}\\
&0&&0&\\
&&1&&
\end{array} \ .
\end{equation}
It is then easy to show that for sufficiently large $N$, we have
\begin{equation}
	P_{t,\bar t}(\mathrm{K3}^N/S_N) = 1 + (t^2 + 21\, t\, \bar t + \bar t^2) + (t^4 + 22\, t^3\, \bar t + 254\, t^2\, \bar t^2 + 
 22\, t\, \bar t^3 + \bar t^4)+\cdots \ .
\end{equation}
In particular, there are $21$ representations of the form $(1,1)_S$. In the following sections, we will compute the $\mathcal{W}^s_\infty$ 
representations associated to these $21$ chiral states, and identify which of them belong to the perturbative Vasiliev sector.
We shall start out, more generally, by determining the coset representations of all states with $\bar{h}=\frac{1}{2}$. 

\subsection{The untwisted sector}\label{sec:4.2}

Let us begin by analysing the $\bar{h}=\frac{1}{2}$ states coming form the untwisted sector of the orbifold \eqref{eq:t4modsnzn}. They 
arise from pairing left- and right-moving states that are separately \emph{not} invariant under $S_{N}\ltimes \mathbb{Z}_2^{N}$, but 
which form a singlet of $S_{N}\ltimes \mathbb{Z}_2^{N}$ when paired together; for example, the simplest such states are of the form
\begin{align}
	\sum_{i=1}^N \psi^{i (\alpha)}\,  \tilde \psi^{\ast i (\beta)}\ ,
\end{align} 
where $\psi^{i (\alpha)}$ and $\tilde \psi^{\ast i (\beta)}$ represent any of the left- or right-moving fermions,  respectively.

In order to count the relevant states we look at the terms proportional to $\bar{q}^{1/2}$ in the 
untwisted sector of the symmetric orbifold of K3 that do not come from the vacuum sector (which describes
states that are separately invariant).\footnote{We should stress that both the twisted and untwisted sector of the 
K3 partition function will contribute to this  $\bar{q}^{1/2}$ order. Here we describe the states that arise in the untwisted 
sector --- the twisted sector contributions will be described in the following subsections.} From the DMVV formula
\cite{Dijkgraaf:1996xw}, we get
\begin{align}
\nonumber \mathcal{Z}^{\rm U}_{1/2}&=\sqrt{q} \left(2 y+\frac{2}{y}\right)+4 q+q^{3/2} \left(2 y^3+\frac{2}{y^3}+14 y
+\frac{14}{y}\right)+q^2 \left(24 y^2+\frac{24}{y^2}+68\right)\\
\nn &+q^{5/2} \left(2 y^5+\frac{2}{y^5}+24 y^3+\frac{24}{y^3}+160 y+\frac{160}{y}\right)\\
 &+q^3 \left(24 y^4+\frac{24}{y^4}+264 y^2+\frac{264}{y^2}+604\right)+\mathcal{O}(q^{7/2})\ ,\label{nontriut}
\end{align}
which can be decomposed into coset representations as 
\begin{align}
\nonumber \mathcal{Z}^{\rm U}_{1/2}&=\chi_{ (0, [1,0,\ldots,0,0])} + \chi_{ (0, [0,0,\ldots,0,1])}+2 \chi_{ (0, [3,0,\ldots,0,0])}
+2 \chi_{ (0, [0,0,\ldots,0,3])}\\
\nonumber & +\chi_{ (0, [1,1,0,\ldots,0,0])}+\chi_{ (0, [0,0,\ldots,0,1,1])}+2 \chi_{ (0, [2,0,\ldots,0,1])}+2 \chi_{ (0, [1,0,\ldots,0,2])}\\
\nonumber &+4 \chi_{ (0, [5,0,\ldots,0,0])}+4 \chi_{ (0, [0,0,\ldots,0,5])}+3 \chi_{ (0, [3,1,0,\ldots,0,0])}+3 \chi_{ (0, [0,0,\ldots,0,1,3])}\\
\nonumber &+2 \chi_{ (0, [1,2,0,\ldots,0,0])}+2 \chi_{ (0, [0,0,\ldots,0,2,1])}+\chi_{ (0, [0,1,1,0,\ldots,0,0])}+\chi_{ (0, [0,0,\ldots,0,1,1,0])}\\
\nonumber &+5 \chi_{ (0, [4,0,\ldots,0,1])}+5 \chi_{ (0, [1,0,\ldots,0,4])}+2 \chi_{ (0, [2,1,0,\ldots,0,1])}+2 \chi_{ (0, [1,0,\ldots,0,1,2])}\\
\nonumber &+2 \chi_{ (0, [0,2,0,\ldots,0,1])}+2 \chi_{ (0, [1,0,\ldots,0,2,0])}+6 \chi_{ (0, [3,0,\ldots,0,2])}+6 \chi_{ (0, [2,0,\ldots,0,3])}\\
 &+\chi_{ (0, [3,0,\ldots,0,1,0])}+\chi_{ (0, [0,1,0,\ldots,0,3])}+3 \chi_{ (0, [1,1,0,\ldots,0,2])}+3 \chi_{ (0, [2,0,\ldots,0,1,1])}+\mathcal{O}(q^{7/2})\ .
\end{align}
As before, the multiplicities are determined by the embedding $S_{N}\ltimes \mathbb{Z}_2^{N} \hookrightarrow \mathrm{U}(N)$, but 
now we consider the multiplicity of the $N$ representation of $S_{N}\ltimes \mathbb{Z}_2^{N}$ (which is the representation
in which the $\tilde{\psi}^{i\beta}$ transform), instead of the trivial representation. 

There are $4$ BPS states of the form $(1,1)_S$  that arise in this sector, and that come from the leading terms in 
$|\mathcal{Z}^{\rm U}_{1/2}|^2$; they are 
\begin{equation}\label{BPSU}
	(0, \mathrm{f}) \otimes \overline{(0,\mathrm{f})}~, \qquad (0, \mathrm{f}) \otimes \overline{(0, \mathrm{f}^*)}~, \qquad (0, \mathrm{f}^*)
	 \otimes \overline{(0, \mathrm{f})}~, \qquad (0, \mathrm{f}^*) \otimes \overline{(0, \mathrm{f}^*)}\ ,
\end{equation}
where $\mathrm{f} = [1,0,\ldots,0,0]$ and $\mathrm{f}^* = [0,0,\ldots,0,1]$ are the fundamental and anti-fundamental 
representations of U($N$), respectively. Note that only two of them, namely 
\begin{equation}\label{BPSVas}
(0, \mathrm{f}) \otimes \overline{(0, \mathrm{f}^*)} \qquad \hbox{and} \qquad 
(0, \mathrm{f}^*) \otimes \overline{(0, \mathrm{f})}
\end{equation}
belong to the perturbative Vasiliev sector, see eq.~\eqref{eq:hpert}.

\subsection{The twisted sectors of $(\mathbb{T}^4)^N \Big/ (S_{N}\ltimes \mathbb{Z}_2^{N})$}\label{sec:4.3}

We now discuss the low-lying states arising from the twisted sectors of \eqref{eq:t4modsnzn}. Since we need to use 
various elements and subgroups of $S_{N}\ltimes \mathbb{Z}_2^{N}$, it is convenient to fix some notation.

The elements of $S_{N}\ltimes \mathbb{Z}_2^{N}$ can be uniquely characterised by a collection of signs,
$s = (s_1,s_2,\ldots,s_N)$, $s_i = \pm$, \emph{followed} by an arbitrary permutation $\pi$. Consequently, we will use the 
notation $\pi_{s}$ for the elements of this group. To avoid clutter, we will omit from the notation the last consecutive $+$ signs, 
e.g., $(-,+,+,\ldots,+) = (-)$. The trivial permutation will be denoted by $\mathbf{1}$.

The twisted sectors of \eqref{eq:t4modsnzn} are labeled by conjugacy classes of $S_{N}\ltimes \mathbb{Z}_2^{N}$. We are 
particularly interested in sectors that contain states with $\bar{h}=1/2$. We will see in 
the following that the relevant conjugacy classes are
\begin{align}\label{conj}
	[\mathbf{1}_{(-)}]\ ,	\qquad [(12)]\ , \qquad [(12)_{(-)}]\ .
\end{align}
We will examine each of them in turn.

\subsubsection{$\mathbb{Z}_2$ torus twist}
Let us consider first the conjugacy class containing the $\mathbb{Z}_2$ element $\mathbf{1}_{(-)}$, i.e., 
the generator of the $\mathbb{Z}_2$ inversion of the underlying $\mathbb{T}^4$. It is easy to see that the 
centraliser subgroup is
\begin{align}
	\label{eq:cz2t}
	C_{\mathbf{1}_{(-)}} = \mathbb{Z}_2 \times (S_{N-1} \ltimes \mathbb{Z}_2^{N-1})\ ,
\end{align}
and that the representation $N$ decomposes as
\begin{align}
	\label{eq:dz2t}
	N \cong 1 \oplus (N-1)\ ,
\end{align}
where the $1$ is odd under the $\mathbb{Z}_2$. Thus the situation is very similar to what was considered in section~7.2
of \cite{Gaberdiel:2014cha}. 

Let us concentrate on the states that transform trivially (separately for left- and right-movers) with respect to the 
$(S_{N-1} \ltimes \mathbb{Z}_2^{N-1})$ factor of the centraliser \eqref{eq:cz2t}, but are either even or odd under the first $\mathbb{Z}_2$. 
Using the explicit transformation properties of the bosonic and fermionic modes under the centraliser, we can compute 
the first few terms of the corresponding characters, and we find
\begin{align}
	\mathcal{Z}^{\rm T}_+ & = q^{1/2}\Big((y+y^{-1}) + 8 q^{1/2} + (y^3+19y+19y^{-1}+y^{-3})q\nonumber\\
	& \phantom{ = \sqrt{q}\big((y+y^{-1}) + 8 q^{1/2} }  + (24y^2 + 112 + 24 y^{-2})q^{3/2} + \cdots \Big)\ ,\\
	\mathcal{Z}^{\rm T}_- & = q^{1/2}\Big(2 + 4(y + y^{-1}) q^{1/2} + (4y^2+32+4y^{-2})q\nonumber\\
	& \phantom{ = q^{1/2}\Big(2 + 4(y + y^{-1}) q^{1/2} }  + (4y^3 + 76y + 76 y^{-1}+4y^{-3})q^{3/2} + \cdots \Big)\ ,	\,
\end{align}
where the sign in $\mathcal{Z}^{\rm T}_\pm$ refers to the eigenvalue under the first $\mathbb{Z}_2$ in \eqref{eq:cz2t}. 
Using similar techniques as in \cite{Gaberdiel:2014cha}, these characters can be decomposed as 
\begin{align}
\nonumber \mathcal{Z}_+^{\rm T} = & \phantom{+} \chi _{ \left([\frac{k}{2},0,\ldots,0],[\frac{k}{2}-1,0,\ldots,0]\right)}
+\chi_{ \left([\frac{k}{2},0,\ldots,0],[\frac{k}{2}+1,0,\ldots,0]\right)}+\chi_{ \left([\frac{k}{2},0,\ldots,0],[\frac{k}{2}+3,0,\ldots,0]\right)}\\
&+\chi_{ \left([\frac{k}{2},0,\ldots,0],[\frac{k}{2}-1,2,0,\ldots,0]\right)}+\chi_{ \left([\frac{k}{2},0,\ldots,0],[\frac{k}{2}+1,0,\ldots,0,2]\right)}
+\mathcal{O}(q^2)\ ,\\
\nonumber \mathcal{Z}_-^{\rm T} = & \phantom{+} \chi_{ \left( [\frac{k}{2},0,\ldots,0], [\frac{k}{2},0,\ldots,0]\right)}
+\chi_{ \left( [\frac{k}{2},0,\ldots,0], [\frac{k}{2}-2,0,\ldots,0]\right)}+\chi_{ \left( [\frac{k}{2},0,\ldots,0], [\frac{k}{2}+2,0,\ldots,0]\right)}\\
\nonumber &+\chi_{ \left( [\frac{k}{2},0,\ldots,0], [\frac{k}{2}+4,0,\ldots,0]\right)}+\chi_{ \left( [\frac{k}{2},0,\ldots,0], [\frac{k}{2},2,0,\ldots,0]\right)}
+\chi_{ \left( [\frac{k}{2},0,\ldots,0], [\frac{k}{2},0,\ldots,0,2]\right)}\\ 
&+\chi_{ \left( [\frac{k}{2},0,\ldots,0], [\frac{k}{2}-2,2,0,\ldots,0]\right)}+\chi_{ \left( [\frac{k}{2},0,\ldots,0], [\frac{k}{2}+2,0,\ldots,0,2]\right)}
 +\mathcal{O}(q^\frac{5}{2}) \ . 
\end{align}
To understand the multiplicities, we can follow the same logic as in \cite{Gaberdiel:2014cha}. First 
we note that the ground state is degenerate due to the fermionic zero modes --- acting with them 
changes only the first entry of the Dynkin label $\Lambda_-=\{k/2+l_0,\Lambda'\}$. The even/odd separation is then 
determined by the index 
$P=(l_0+\sum_i\Lambda'_i)~ \mathrm{mod} ~ 2$, while the multiplicity with which each representation appears in the decomposition equals
the number of $(S_{N-1}\ltimes\mathbb{Z}_2^{N-1})$ singlets contained in the corresponding ${\rm U}(N-1)$ representation.
This then reduces to the same computation as in the untwisted sector.

It is clear from the structure of $\mathcal{Z}_+^{\rm T}$ that the ground states of $|\mathcal{Z}_+^{\rm T}|^2$ describe a BPS state of 
the form $(1,1)_S$.  Since there are $16$ $\mathbb{Z}_2$-twisted sectors in ${\rm K3}= \mathbb{T}^4 / \mathbb{Z}_2$, there are a total 
of $16$ such states that sit in the ${\cal W}^s_\infty$ representations
\begin{equation}
	\label{eq:16chiralpr}
	16 \cdot ([k/2,0,\ldots,0],[k/2-1,0,\ldots,0]) \otimes \overline{([k/2,0,\ldots,0],[k/2-1,0,\ldots,0])}~.
\end{equation}
None of these states belongs to the perturbative Vasiliev sector.

\subsubsection{$S_2$-twisted sector of $\mathrm{K3}^N/S_N$}

Next we consider the conjugacy class $[(12)]$ corresponding to the permutation $(12)$ of the $N$ copies. The centraliser subgroup is 
in this case 
\begin{align}
	C_{(12)} = (S_2 \times \mathbb{Z}_2^{D}) \times (S_{N-2} \ltimes \mathbb{Z}_2^{N-2})\ ,
\end{align}
where $\mathbb{Z}_2^{D}$ is the diagonal subgroup of $\mathbb{Z}_2^2$, and the product $S_2 \times \mathbb{Z}_2^D$ is \emph{not} semidirect. 
The standard representation now breaks into
\begin{align}
  N \cong 1^{(+,-)} \oplus 1^{(-,-)} \oplus (N-2)\ ,
\end{align}
where $s_1$ and $s_2$ in $1^{(s_1,s_2)}$ indicate the charges with respect to $S_2$ and $\mathbb{Z}_2^D$, respectively. 
As before we shall only consider the states that are separately singlets with respect to the $ (S_{N-2} \ltimes \mathbb{Z}_2^{N-2})$ factor of 
the centraliser. Then there are four such sectors, depending on the eigenvalues $(s_1,s_2)$ under the $(S_2 \times \mathbb{Z}_2^{D})$
group. 
Since the fermionic zero-modes have charge $(-,-)$ with respect to both $\mathbb{Z}_2$'s, there will be two $(+,+)$ ground states of 
charge $y$ and $y^{-1}$ respectively, and two $(-,-)$ ground states. By acting with the oscillators on the various groundstates, we get
\begin{align}
	\label{eq:zpp} \mathcal{Z}_{(+,+)}^{\rm T} & = q^{1/2}\Big((y+y^{-1}) + 8 q^{1/2} + (2y^3+24y+24y^{-1}+2y^{-3})q\nonumber\\
	& \phantom{ = \sqrt{q}\big((y+y^{-1}) + 8 q^{1/2} }  + (40y^2 + 160 + 40 y^{-2})q^{3/2} + \cdots \Big)\ ,\\
	\label{eq:zmm} \mathcal{Z}_{(-,-)}^{\rm T} & = q^{1/2}\Big(2 + 4(y + y^{-1}) q^{1/2} + (6y^2+40+6y^{-2})q\\
	& \phantom{ = q^{1/2}\Big(2 + 4(y + y^{-1}) q^{1/2} }  + (8y^3 + 112y + 112 y^{-1}+8y^{-3})q^{3/2} + \cdots \Big)\ , \nonumber \\
	\mathcal{Z}_{(+,-)}^{\rm T} & = (2y^2 + 4 + 2y^{-2}) q + \cdots,\\[2pt]
	\mathcal{Z}_{(-,+)}^{\rm T} & = (4y + 4y^{-1}) q + \cdots~.
\end{align}
In terms of the coset representations, the first two characters then decompose as 
\begin{align}
\nn \mathcal{Z}_{(+,+)}^{\rm T}&=\chi_{ \left([\frac{k}{2},0,\ldots,0 ],[\frac{k}{2}-1,0,\ldots,0 ]\right)}
+\chi_{ \left([\frac{k}{2},0,\ldots,0 ],[\frac{k}{2}+1,0,\ldots,0 ]\right)}\\
\nonumber &+2 \chi_{ \left([\frac{k}{2},0,\ldots,0 ],[\frac{k}{2}-1,2,0,\ldots,0 ]\right)}+2 \chi_{ \left([\frac{k}{2},0,\ldots,0 ],[\frac{k}{2}+1,0,\ldots,0,2 ]\right)}\\
&+\chi_{ \left([\frac{k}{2},0,\ldots,0 ],[\frac{k}{2},1,0,\ldots,0,1 ]\right)}+\chi_{ \left([\frac{k}{2},0,\ldots,0 ],[\frac{k}{2}+3,0,\ldots,0 ]\right)}
+\mathcal{O}(q^{2})\label{cosets2m}\ ,\\
\nn \mathcal{Z}_{(-,-)}^{\rm T}&=\chi_{ \left([\frac{k}{2},0,\ldots,0 ],[\frac{k}{2},0,\ldots,0 ]\right)}
+\chi_{ \left([\frac{k}{2},0,\ldots,0 ],[\frac{k}{2}-2,0,\ldots,0 ]\right)}
+\chi_{ \left([\frac{k}{2},0,\ldots,0 ],[\frac{k}{2}+2,0,\ldots,0 ]\right)}\\
\nonumber &+\chi_{ \left([\frac{k}{2},0,\ldots,0 ],[\frac{k}{2}+4,0,\ldots,0 ]\right)}+2 \chi_{ \left([\frac{k}{2},0,\ldots,0 ],[\frac{k}{2},2,0,\ldots,0 ]\right)}
+2 \chi_{ \left([\frac{k}{2},0,\ldots,0 ],[\frac{k}{2},0,\ldots,0,2 ]\right)}\\
\nonumber &+2 \chi_{ \left([\frac{k}{2},0,\ldots,0 ],[\frac{k}{2}-2,2,0,\ldots,0 ]\right)}+2 \chi_{ \left([\frac{k}{2},0,\ldots,0 ],[\frac{k}{2}+2,0,\ldots,0,2 ]\right)}
+\chi_{ \left([\frac{k}{2},0,\ldots,0 ],[\frac{k}{2}-1,1,0,\ldots,0,1 ]\right)}\\
&+\chi_{ \left([\frac{k}{2},0,\ldots,0 ],[\frac{k}{2}+1,1,0,\ldots,0,1 ]\right)}+\mathcal{O}(q^{5/2})\ .\label{cosets2p}
\end{align} 
The multiplicities can be determined exactly as in the previous subsection, however now the relevant embedding is 
$S_{N-2}\ltimes \mathbb{Z}_2^{N-2}\hookrightarrow {\rm U}(N-1)$. It is then convenient to consider the chain of embeddings 
$S_{N-2}\ltimes \mathbb{Z}_2^{N-2}\hookrightarrow {\rm U}(N-2) \subset {\rm SU}(N-1)$, where the $N-1$ of 
${\rm U}(N-1)$ decomposes into $(N-2)\oplus 1$ of ${\rm U}(N-2)$. As a consequence, besides the 
$S_{N-2}\ltimes \mathbb{Z}_2^{N-2}$ invariant excitations from the $N-2$ part, there will be additional contributions from  the 
${\rm U}(N-2)$ singlet. However, the $\mathbb{Z}_2^D$ projection implies that only the states built from an even number of such singlets 
will survive.

Now, only the sector $|\mathcal{Z}_{(+,+)}|^2$ gives rise to a chiral primary state of the form $(1,1)_S$, whose
${\cal W}^s_\infty$ representation is 
\begin{equation}\label{BPSS2}
	([k/2,0,\ldots,0],[k/2-1,0,\ldots,0]) \otimes \overline{([k/2,0,\ldots,0],[k/2-1,0,\ldots,0])}\ .
\end{equation}
This is of the same form as the states in \eqref{eq:16chiralpr}. In particular, this BPS state does not appear in the
perturbative part of the Vasiliev theory. 

\subsubsection{$S_2/\mathbb{Z}_2$ twisted sector}

Finally, we examine the case where the twist is in the conjugacy class $[(12)_{(-)}]$. The centraliser is then 
\begin{align}
  C_{(12)_{(-)}} = \mathbb{Z}_4 \times (S_{N-2} \ltimes \mathbb{Z}_2^{N-2})\ ,
\end{align}
where the group $\mathbb{Z}_4$ is generated by the element $(12)_{(-)}$.
The standard representation breaks now into
\begin{align}
  N \cong 1^{(i)} \oplus 1^{(-i)} \oplus (N-2)\ ,
\end{align}
where $1^{(i)}$ and $1^{(-i)}$ have eigenvalue $i$ and $-i$ under the element $(12)_{(-)}$ respectively. The
corresponding fields will then have moding $\frac14 + n$ and $-\frac14 + n=\frac{3}{4}+n'$ in this twisted sector, 
respectively. In particular, the ground state is non-degenerate, and its ground-state energy equals
\begin{align}
h = |\frac14|+|-\frac14| = \frac12\ ,
\end{align}
thus potentially allowing for a BPS state $(1,1)_S$. We will only consider states that are separately (for left- and 
right-movers) a singlet under 
the $(S_{N-2} \ltimes \mathbb{Z}_2^{N-2})$ factor of the centraliser; there are then four sectors $\mathcal{Z}_P$, 
$P=0,1,2,3$, where
$\mathcal{Z}_P$ has eigenvalue $i^P$ under the $\mathbb{Z}_4$ generator of $(12)_{(-)}$. Expanding out the 
relevant characters we get 
\begin{align}
	\label{eq:z0} \mathcal{Z}_{0}^{\rm T} & = q^{1/2} \Big(1+(8y+8y^{-1})q^{1/2} + (15y^2 + 104 + 15y^{-2})q \\
	& \phantom{ = q^{1/2} \Big(1+(8y+8y^{-1})q^{1/2} }  + (16y^3 + 392y + 392y^{-1} + 16y^{-3})q^{3/2} + \cdots \Big)\ ,  \nonumber \\
	\mathcal{Z}_{1}^{\rm T} & = (2y+2y^{-1})q^{3/4} + (4y^2 + 40 + 4y^{-2})q^{5/4} \\
	 &\phantom{ = (2y+2y^{-1})q^{3/4} + (4y^2 + 40 } +(4y^3 + 164 y + 164 y^{-1} + 4 y^{-3})q^{7/4}+\cdots\ ,  \nonumber \\
	\mathcal{Z}_{2}^{\rm T} & = (y^2+14+y^{-2})q + (64y + 64 y^{-1})q^{3/2}\\
	 &\phantom{ = (y^2+14+y^{-2})q + (64y} + (y^4 + 118 y^2 + 562 + 118 y^{-2} + y^{-4})q^{2}+\cdots\ , \nonumber \\
	\mathcal{Z}_{3} ^{\rm T}& = 4q^{3/4} + (24y + 24y^{-1})q^{5/4} + (44y^2 + 248 + 44 y^{-2})q^{7/4}+\cdots\ . \label{eq:z3}
\end{align}
We can also write these characters in terms of coset representations, and the explicit formulae are given in the appendix. 
Note that none of these sectors contains a $(1,1)_S$ BPS state --- indeed, only ${\cal Z}_0$ has 
$h=\frac{1}{2}$, but the corresponding ground state is a singlet, and does not transform in the $j=\frac{1}{2}$ 
of the R-symmetry ${\rm SU}(2)$. 

\subsection{A consistency check}

At the beginning of the section we argued that the states with $\bar{h}=1/2$ are accounted for by the 
untwisted sector, as well as the twisted sectors corresponding to the conjugacy classes (\ref{conj}). We 
can now check this by comparing the full partition function, as derived by the DMVV formula \cite{Dijkgraaf:1996xw}, 
with the ansatz
\begin{align}
\mathcal{Z}_{{\rm ansatz}} = \left| \mathcal{Z}_{\rm vac} \right|^2 + \left| \mathcal{Z}^{\rm U}_{1/2} \right|^2 
+ 16 \Bigl( \left| \mathcal{Z}^{\rm T}_{+} \right|^2  + \left| \mathcal{Z}^{\rm T}_{-} \right|^2 \Bigr) 
+ \left| \mathcal{Z}^{\rm T}_{(+,+)} \right|^2  + \left| \mathcal{Z}^{\rm T}_{(-,-)} \right|^2
+ \left| \mathcal{Z}^{\rm T}_{0} \right|^2   \ . 
\end{align}
We have checked that this ansatz reproduces correctly all terms of order $\bar{q}^{1/2}$ with arbitrary powers
of $q$ (up to the order to which we have worked out these characters, i.e., up to order $q^{3/2}$). This confirms
that we have accounted correctly for all terms with $\bar{h}=1/2$. In particular, we have also found all $21$ 
$(1,1)_S$ BPS states: $4$ come from the untwisted sector, see eq.~\eqref{BPSU}, while $16$ come from the 
$\mathbb{Z}_2$ twisted sector of the torus orbifold, see eq.~\eqref{eq:16chiralpr}, and the last one arises in the 
$(12)$-twisted sector, see eq.~\eqref{BPSS2}. Of these $21$ BPS states, only two are present in the perturbative
Vasiliev theory, namely two of the $4$ states in eq.~\eqref{BPSU}. Note that all of the $17$ BPS states arising from
the twisted sector sit in the same coset representation, see eqs.~\eqref{eq:16chiralpr} and \eqref{BPSS2}.

\subsection{More general BPS states}

In the previous subsection we identified which of the $(1,1)_S$ BPS states of the symmetric orbifold are accounted
for by the perturbative Vasiliev theory. In this section we want to analyse this question for more general BPS states.
It is not difficult to show that the BPS states that appear in the perturbative Vasiliev spectrum \eqref{eq:hpert} are of the form 
\begin{align}
(0,[0^{n_1-1},1,0^{N-n_1-n_2-1},1,0^{n_2-1}])\otimes \overline{(0,[0^{n_2-1},1,0^{N-n_1-n_2-1},1,0^{n_1-1}])}\ ,\label{gensym}
\end{align} 
where $n_1,n_2=0,1,2,\ldots$, and not both $n_1$ and $n_2=0$. (Note that $n_1=0$, $n_2=1$, for example, corresponds to the 
representation $(0,{\rm f}^{\ast}) \otimes \overline{(0;{\rm f})}$.) They describe BPS states associated to 
\begin{equation}
(n_1 + n_2, n_1 + n_2)_S \ . 
\end{equation}
However, except for the cases $(n_1,n_2)=(1,0)$ and  $(n_1,n_2)=(0,1)$, whose top components correspond to 
\begin{equation}
\sum_{i=1}^N \psi^{i (1)} \tilde \psi^{\ast i(2)}\ , \qquad
\sum_{i=1}^N \psi^{\ast i (2)} \tilde \psi^{i(1)}\ , 
\end{equation}
respectively,  they are all multi-particle, i.e., they involve more than one sum over $i$, as dictated 
by the fermionic statistics.  Thus 
the only single-particle BPS states of the Vasiliev theory are two states with $(1,1)_S$, corresponding
to $n_1+n_2=1$. Thus the perturbative Vasiliev
theory only captures a tiny part of the BPS spectrum of the full string background.

\section{Conclusions}

In this paper we have studied how a slightly modified version of the ${\cal N}=4$ higher spin -- CFT duality of 
\cite{Gaberdiel:2013vva} can be naturally realised as a subsector of the symmetric orbifold of ${\rm K3}=\mathbb{T}^4/\mathbb{Z}_2$,
which in turn is believed to be dual to string theory on ${\rm AdS}_3 \times {\rm S}^3 \times {\rm K3}$ in the 
tensionless limit. Most of the analysis was quite parallel to what was done in \cite{Gaberdiel:2014cha}, but there
were also important differences: the relevant ${\cal W}_\infty$ algebra is not directly a Wolf space coset, but is obtained
in the limit $k\rightarrow \infty$ upon restricting to a consistent subalgebra, see eq.~(\ref{dualitys}). As a consequence, the structure of the
branching rules (that determine the multiplicities of the corresponding ${\cal W}_\infty^s$ representations)
was somewhat different to what appeared in  \cite{Gaberdiel:2014cha}. Similarly, the structure of the twisted sectors
of the symmetric orbifold of ${\rm K3}=\mathbb{T}^4/\mathbb{Z}_2$ is quite rich, and we have identified all of the
low-lying twisted sector states in terms of ${\cal W}_\infty^s$ representations, see sections~\ref{sec:4.2} and
\ref{sec:4.3}. In particular, this allowed us to analyse which of the
chiral primaries of the symmetric orbifold are actually contained in the perturbative higher spin theory, and we found
that this is only true for a tiny number of single-particle states.

It would be interesting to understand the structure of the stringy symmetry algebra for this case; since all symmetry 
generators come from the untwisted sector, the relevant algebra should be a subalgebra of the stringy algebra for the
$\mathbb{T}^4$ case, whose structure was studied in \cite{Gaberdiel:2015mra}. It would also be interesting to study
the behaviour of the symmetry currents under the perturbation that corresponds to switching on the tension; again,
because of the same reason, this should be very similar to the corresponding analysis for the $\mathbb{T}^4$ 
case, see \cite{GCZ}.\footnote{For the ${\cal N}=3$ case that was proposed in \cite{Creutzig:2014ula}, 
a similar analysis was recently performed  in  \cite{Hikida:2015nfa}.} In this paper we have studied the case of the K3 being 
described by a $\mathbb{Z}_2$ torus orbifold,
but there are other K3 sigma models that have an explicit CFT description, e.g., as Gepner models or as other torus orbifolds,
see \cite{Nahm:1999ps}. 
It would be interesting to repeat the above analysis for these models to see in which way the symmetry algebra depends on the
point in the K3 moduli space. Finally, it would be very interesting more generally to study implications of the gigantic stringy symmetry 
for various aspects of string theory.

\section*{Acknowledgements}

We thank Constantin Candu and Rajesh Gopakumar for useful discussions. C.P.\ greatly appreciates the warm hospitality of the 
Korea Institute for Advanced Study and the University of Michigan, Ann Arbor during various stages of this work. 
The work of C.P.\ is supported by a grant of the Swiss National Science Foundation. 
We also gratefully acknowledge the support by the NCCR SwissMAP.

\appendix

\section{The $S_2/\mathbb{Z}_2$ twisted sector characters}

In this appendix we write the characters from the $S_2/\mathbb{Z}_2$ twisted sector, see eqs.\ (\ref{eq:z0}) -- (\ref{eq:z3}), in 
terms of coset representations. We find 
{
\begin{align}
\nonumber 
\mathcal{Z}_{0}^{\rm T}=~&\chi_{ \left( [\frac{k}{4},0,\ldots,0,\frac{k}{4}], [\frac{k}{4},0,\ldots,0,\frac{k}{4}]\right)}
+\chi_{ \left( [\frac{k}{4},0,\ldots,0,\frac{k}{4}], [\frac{k}{4},0,\ldots,0,\frac{k}{4}-4]\right)}
+\chi_{ \left( [\frac{k}{4},0,\ldots,0,\frac{k}{4}], [\frac{k}{4}-4,0,\ldots,0,\frac{k}{4}]\right)}\\
\nonumber &+\chi_{ \left( [\frac{k}{4},0,\ldots,0,\frac{k}{4}], [\frac{k}{4}-3,0,\ldots,0,\frac{k}{4}-1]\right)}
+\chi_{ \left( [\frac{k}{4},0,\ldots,0,\frac{k}{4}], [\frac{k}{4}-2,0,\ldots,0,\frac{k}{4}-2]\right)}\\
\nonumber &+\chi_{ \left( [\frac{k}{4},0,\ldots,0,\frac{k}{4}], [\frac{k}{4}-1,0,\ldots,0,\frac{k}{4}-3]\right)}
+\chi_{ \left( [\frac{k}{4},0,\ldots,0,\frac{k}{4}], [\frac{k}{4}-1,0,\ldots,0,\frac{k}{4}+1]\right)}\\
\nonumber &+\chi_{ \left( [\frac{k}{4},0,\ldots,0,\frac{k}{4}], [\frac{k}{4}+2,0,\ldots,0,\frac{k}{4}-2]\right)}
+\chi_{ \left( [\frac{k}{4},0,\ldots,0,\frac{k}{4}], [\frac{k}{4}+2,0,\ldots,0,\frac{k}{4}+2]\right)}\\
\nonumber &+\chi_{ \left( [\frac{k}{4},0,\ldots,0,\frac{k}{4}], [\frac{k}{4}-2,0,\ldots,0,\frac{k}{4}+2]\right)}
+\chi_{ \left( [\frac{k}{4},0,\ldots,0,\frac{k}{4}], [\frac{k}{4}+1,0,\ldots,0,\frac{k}{4}-1]\right)}\\
 &+\chi_{ \left( [\frac{k}{4},0,\ldots,0,\frac{k}{4}], [\frac{k}{4}-2,2,0,\ldots,0,\frac{k}{4}]\right)}
 +\chi_{ \left( [\frac{k}{4},0,\ldots,0,\frac{k}{4}], [\frac{k}{4},0,\ldots,0,2,\frac{k}{4}-2]\right)}+\mathcal{O}(q^{2}) \ , \label{coset04} \\
\nonumber \mathcal{Z}_{1}^{\rm T}=~&\chi_{ \left( [\frac{k}{4},0,\ldots,0,\frac{k}{4}], [\frac{k}{4},0,\ldots,0,\frac{k}{4}-3]\right)}
+\chi_{ \left( [\frac{k}{4},0,\ldots,0,\frac{k}{4}], [\frac{k}{4},0,\ldots,0,\frac{k}{4}+1]\right)}\\
\nonumber &+\chi_{ \left( [\frac{k}{4},0,\ldots,0,\frac{k}{4}], [\frac{k}{4}-4,0,\ldots,0,\frac{k}{4}+1]\right)}
+\chi_{ \left( [\frac{k}{4},0,\ldots,0,\frac{k}{4}], [\frac{k}{4}-3,0,\ldots,0,\frac{k}{4}]\right)}\\
\nonumber &+\chi_{ \left( [\frac{k}{4},0,\ldots,0,\frac{k}{4}], [\frac{k}{4}-2,0,\ldots,0,\frac{k}{4}-1]\right)}
+\chi_{ \left( [\frac{k}{4},0,\ldots,0,\frac{k}{4}], [\frac{k}{4}-1,0,\ldots,0,\frac{k}{4}-2]\right)}\\
\nonumber &+\chi_{ \left( [\frac{k}{4},0,\ldots,0,\frac{k}{4}], [\frac{k}{4}-1,0,\ldots,0,\frac{k}{4}+2]\right)}
+\chi_{ \left( [\frac{k}{4},0,\ldots,0,\frac{k}{4}], [\frac{k}{4}+1,0,\ldots,0,\frac{k}{4}]\right)}\\
\nonumber &+\chi_{ \left( [\frac{k}{4},0,\ldots,0,\frac{k}{4}], [\frac{k}{4}+1,0,\ldots,0,\frac{k}{4}-4]\right)}
+\chi_{ \left( [\frac{k}{4},0,\ldots,0,\frac{k}{4}], [\frac{k}{4}+2,0,\ldots,0,\frac{k}{4}-1]\right)}\\
\nonumber &+\chi_{ \left( [\frac{k}{4},0,\ldots,0,\frac{k}{4}], [\frac{k}{4},0,\ldots,0,2,\frac{k}{4}-1]\right)}
+\chi_{ \left( [\frac{k}{4},0,\ldots,0,\frac{k}{4}], [\frac{k}{4}-2,2,0,\ldots,0,\frac{k}{4}+1]\right)}\\
 &+\chi_{ \left( [\frac{k}{4},0,\ldots,0,\frac{k}{4}], [\frac{k}{4}-1,2,0,\ldots,0,\frac{k}{4}]\right)}
 +\chi_{ \left( [\frac{k}{4},0,\ldots,0,\frac{k}{4}], [\frac{k}{4}+1,0,\ldots,0,2,\frac{k}{4}-2]\right)}+\mathcal{O}(q^{9/4})\ ,\label{coset14}\\
\nonumber\mathcal{Z}_{2}^{\rm T}=~&\chi_{ \left( [\frac{k}{4},0,\ldots,0,\frac{k}{4}], [\frac{k}{4},0,\ldots,0,\frac{k}{4}-2]\right)}
+\chi_{ \left( [\frac{k}{4},0,\ldots,0,\frac{k}{4}], [\frac{k}{4},0,\ldots,0,\frac{k}{4}+2]\right)}\\
\nonumber & +\chi_{ \left( [\frac{k}{4},0,\ldots,0,\frac{k}{4}], [\frac{k}{4}-3,0,\ldots,0,\frac{k}{4}+1]\right)}
+\chi_{ \left( [\frac{k}{4},0,\ldots,0,\frac{k}{4}], [\frac{k}{4}-2,0,\ldots,0,\frac{k}{4}]\right)}\\
\nonumber &+\chi_{ \left( [\frac{k}{4},0,\ldots,0,\frac{k}{4}], [\frac{k}{4}-1,0,\ldots,0,\frac{k}{4}-1]\right)}
+\chi_{ \left( [\frac{k}{4},0,\ldots,0,\frac{k}{4}], [\frac{k}{4}+1,0,\ldots,0,\frac{k}{4}-3]\right)}\\
 &+\chi_{ \left( [\frac{k}{4},0,\ldots,0,\frac{k}{4}], [\frac{k}{4}+1,0,\ldots,0,\frac{k}{4}+1]\right)}
 +\chi_{ \left( [\frac{k}{4},0,\ldots,0,\frac{k}{4}], [\frac{k}{4}+2,0,\ldots,0,\frac{k}{4}]\right)}+\mathcal{O}(q^{2})\ ,\label{coset24}\\
\nonumber \mathcal{Z}_{3}^{\rm T}=~&\chi_{ \left( [\frac{k}{4},0,\ldots,0,\frac{k}{4}], [\frac{k}{4},0,\ldots,0,\frac{k}{4}-1]\right)}
+\chi_{ \left( [\frac{k}{4},0,\ldots,0,\frac{k}{4}], [\frac{k}{4}-2,0,\ldots,0,\frac{k}{4}+1]\right)}\\
\nonumber & +\chi_{ \left( [\frac{k}{4},0,\ldots,0,\frac{k}{4}], [\frac{k}{4}-1,0,\ldots,0,\frac{k}{4}]\right)}
+\chi_{ \left( [\frac{k}{4},0,\ldots,0,\frac{k}{4}], [\frac{k}{4}+1,0,\ldots,0,\frac{k}{4}-2]\right)}\\
 &+\chi_{ \left( [\frac{k}{4},0,\ldots,0,\frac{k}{4}], [\frac{k}{4}+1,0,\ldots,0,\frac{k}{4}+2]\right)}
 +\chi_{ \left( [\frac{k}{4},0,\ldots,0,\frac{k}{4}], [\frac{k}{4}+2,0,\ldots,0,\frac{k}{4}+1]\right)}+\mathcal{O}(q^{7/4})\ .\label{coset34}
\end{align}
Note that the ground state, $\left( [\frac{k}{4},0,\ldots,0,\frac{k}{4}], [\frac{k}{4},0,\ldots,0,\frac{k}{4}]\right)$, can
be determined by the procedure outlined in \cite{Gaberdiel:2014cha, Gaberdiel:2014vca}, with the relevant twist 
being $\alpha = (\frac{1}{4},0,\ldots,0,-\frac{1}{4})$.
The coset representations $\left( [\frac{k}{4},0,\ldots,0,\frac{k}{4}], [\frac{k}{4}+l_1,\Lambda',\frac{k}{4}+l_{N-1}]\right)$ that 
contribute in each sector are constrained by the selection rule
\begin{align}
P&=l_1+l_{N-1}+\sum_i \Lambda '_i \quad (\hbox{mod $4$})   \ ,\label{proj4}
\end{align}
while the multiplicities of $\Lambda'$ are determined in the usual way via the embedding 
$S_{N-2}\ltimes \mathbb{Z}_2^{N-2} \hookrightarrow {\rm U}(N-2)$.

\bibliographystyle{utphys}
\bibliography{k3_notes}

\end{document}